\documentclass[12pt]{spieman}  % 12pt font required by SPIE;
\usepackage{amsmath,amsfonts,amssymb}
\usepackage{graphicx}
\usepackage{setspace}
\usepackage{tocloft}
\usepackage{bm}
\usepackage{makecell}
\usepackage{verbatimbox}
\usepackage[skip=1pt]{caption}
\title{Removing Multi-frame Gaussian Noise by 
Combining Patch-based Filters with Optical Flow}

\author[a,*]{Kireeti Bodduna}
\author[a,*]{Joachim Weickert}
\affil[a]{Mathematical Image Analysis Group, 
Faculty of Mathematics and Computer Science, 
Saarland University, 66041 Saarbr{\"u}cken, Germany.}

\cftpagenumbersoff{figure}
\cftpagenumbersoff{table} 
\begin{document} 
\maketitle

\begin{abstract}
Patch-based approaches such as 3D block matching (BM3D) and 
non-local Bayes (NLB) are widely accepted filters
for removing Gaussian noise from single-frame images.
In this work, we propose three extensions 
for these filters when there exist multiple frames of the 
same scene. The first of them employs reference 
patches on every frame instead of a commonly used 
single reference frame method, thus utilizing  
the complete available information. 
The remaining two techniques use a separable spatio-temporal 
filter to reduce interactions between
dissimilar regions, hence mitigating artifacts.
In order to deal with non-registered datasets 
we combine all our extensions with robust optical flow computation. 
Two of our proposed multi-frame filters 
outperform existing extensions on most occasions by a 
significant margin while also being competitive with a 
state-of-the-art neural network-based technique. 
Moreover, one of these two strategies is the fastest among all 
due to its separable design. 
\end{abstract}

% Include a list of up to six keywords after the abstract

% Include email contact information for corresponding author
{\noindent \footnotesize\textbf{*} 
\linkable{\{bodduna,weickert\}@mia.uni-saarland.de} }

\begin{spacing}{1}   % use double spacing for rest of manuscript

\section{Introduction}
\label{sec:1}
Restoring images corrupted with various types of noise degradations     
is a classical image processing problem. 
Additive white Gaussian noise (AWGN), Poissonian and
mixture noise types are the most studied
noise models. 
AWGN elimination is particularly 
important because it combined with variance stabilizing 
transformations \cite{Anscombe1948PoissonTrans,
makitalo2011optimal,makitalo2013optimal} for also removing the latter 
two types of noise. 

In the single-frame AWGN elimination scenario 
\cite{BCM05, BCM05a, DFKE2007, Lebrun2012, 
LBM2013, LBM2013a}, 
BM3D \cite{DFKE2007, Lebrun2012} 
and NLB \cite{LBM2013, LBM2013a}  
produce superior results. 
{Early contributions to Bayesian non-local denoising can be 
found in the works of Awate and Whitaker
\cite{awate2005nonparametric, awate2007feature}.
For a comprehensive survey on image filtering, we refer to 
Milanfar \cite{milanfar2012tour}}.
BM3D and NLB are non-local 
patch-based methods which utilize the similar information 
available at distant regions in the image. More precisely, 
they filter a 3D group of similar patches.
BM3D in particular is a quasi-standard for modern denoising
algorithms. It is used as a benchmark in articles that involve 
both neural network-based techniques \cite{cruz2018nonlocality} and 
traditional approaches \cite{LBM2013}. 

Multi-frame filters 
\cite{luisier2009MultiFrame, delpretti2008multiframe, 
hasan2018denoising, boulanger2010patch, dong2015low, 
dong2018robust, hao2018patch, fang2012sparsity, 
rodrigo2011SparsityMultiFrame, zhang2009MultiFrame, 
buades2005denoising, tico2008multi,
maggioni2012nonlocal, maggioni2012video, 
arias2018video, Buades2009, buades2010MultiFrame,
buades2016patch, dabov2007video}, on the other hand, 
utilize information from multiple frames of the same scene to 
compute the final denoised image. In this work, we concentrate on 
the fundamental problem of finding 
general approaches that can optimally extend single-frame 
patch-based methods such as NLB and BM3D to the multi-frame scenario. 

There already exist two types of extensions  
\cite{buades2005denoising,  
tico2008multi, maggioni2012nonlocal, maggioni2012video, 
arias2018video} for BM3D and NLB. 
Methods from the first category search for similar 
2D patches from all the available frames. 
However, they use just one reference frame
for filtering purposes, thus making limited use of the 
available information \cite{buades2005denoising, tico2008multi}.
Extensions from the other category take 
privilege of having more data in 3D spatio-temporal patches 
\cite{maggioni2012video, maggioni2012nonlocal, arias2018video}. 
Nevertheless, techniques which utilize 2D patches on 
multiple reference frames   
and those which seperately filter information in the 
spatial and temporal dimensions, have not been studied. 
The latter can reduce undesirable interactions 
between regions of dissimilar greyvalues. 
Furthermore, a careful and 
systematic evaluation of these extensions is also missing. 

{\bf Our Contribution.} In order to address the above problems, 
in our recent conference paper \cite{bodduna2019MultiFrame} 
we introduced three extensions which can be divided into two 
categories: Firstly, we employed the 2D patch similarity approach 
of Buades et al. \cite{buades2005denoising} and Tico
\cite{tico2008multi} but using every frame as a reference one 
for filtering purposes. This ensured that we made use of the 
complete available information. Secondly, we introduced two other
extensions which benefit from separately filtering
the different types of data in the temporal 
and the spatial dimensions. The first one 
performs a simple temporal averaging followed by 
a single-frame spatial filtering, while the other  
reverses this order. 

In the present work we additionally introduce three novel contributions:  
Firstly, we also consider non-registered data. In contrast to 
our conference work \cite{bodduna2019MultiFrame}, we combine our 
multi-frame filters with robust optical flow methods for dealing 
with the inter-frame motion. Such a study is really interesting as 
the utilisation of motion compensation was avoided by 
Arias and Morel \cite{arias2018video} for circumventing 
motion estimation errors.
In fact, contrary to most works on multi-frame denoising,
we juxtapose the filter performance simultaneously 
for perfectly registered and for non-registered data.
For the latter scenario, we pay special attention to parameter 
optimisation of the optical flow approaches. Such an analysis 
provides valuable additional insights into the 
importance of well optimized motion estimation in multi-frame 
denoising.  

Secondly, we provide the first comprehensive evaluation of general 
strategies how to extend single-frame filters to multi-frame ones.
In our previous work \cite{bodduna2019MultiFrame} we 
applied all the proposed extensions to just BM3D. In this paper,  
we also include the NLB denoising filter. Our  
evaluations include very high AWGN noise levels. 
Such large amplitudes of noise, which are 
consistenly ignored in the literature, are very relevant for 
microscopic and medical imaging applications.

Last but not least, we propose better parameter selection strategies
for our filters than in our conference paper. 
We shall see that this will even change
the order in our experimental rankings. For the sake of 
completeness, we also include three state-of-the-art 
multi-frame denoising solutions 
\cite{ maggioni2012video, arias2018video, davy2019non} 
in the evaluation part, which was missing in our preceding
paper \cite{bodduna2019MultiFrame}. 
The neural network-based approach presented in 
\cite{davy2019non, davy2020video} is one among the many
learning-based multi-frame filtering strategies \cite{godard2018deep, mildenhall2018burst} adopted nowadays.

\par {\bf Paper Structure.} 
In Section \ref{sec:modelling} we first review the central ideas 
behind the design of NLB and BM3D filters. We then introduce the 
five multi-frame extensions including our proposed techniques, along 
with the existing robust optical flow methods employed for 
registration. In the ensuing Section \ref{sec:exp}, the new 
optimal parameter selections for our extensions are 
presented. We also 
showcase the results of several denoising experiments along
with detailed explanations behind 
the observed ranking of various techniques. Finally, 
in Section \ref{sec:conclusion} we conclude our work with a summary 
and an outlook. 
 
% -----------------------------------------------------------%
\section{Modeling and Theory}
\label{sec:modelling}
% -----------------------------------------------------------%
\subsection{Filters for Single-frame Image Datasets}
\label{sec:single_frame}
NLB \cite{LBM2013, LBM2013a} and BM3D \cite{DFKE2007, Lebrun2012} 
are non-local patch-based denoising methods which consider  
similar information from distant regions in the image.
Both single-frame filters are two step approaches which 
combine the denoised image of the initial step with the noisy 
image in order to derive the final noise-free image. Furthermore, 
both of these steps are split into three sub-steps each,
namely grouping, collaborative filtering and aggregation. \\\\
\textit{Grouping:} 
In order to exploit the advantage of having more information,
for every noisy reference patch considered,
one forms a 3D group of similar patches using $L_2$ distance. \\\\
\textit{Collaborative Filtering:} 
The term "collaborative" has a 
literal meaning here: Each patch in a group collaborates 
with the rest of them for simultaneous and efficient filtering. 
In NLB, one uses Bayesian filtering (in both 
main steps) to denoise the 3D groups. 
In BM3D, a hard thresholding (first main step) and 
Wiener filtering (second main step) are employed. \\\\
\textit{Aggregation:} In order to derive the final denoised image, 
one computes a weighted averaging of the several denoised versions 
of every pixel. 
% -----------------------------------------------------------%
\subsection{Multi-frame Extensions of Single-frame Filters}
In this section, we describe five multi-frame extensions 
for the above mentioned single-frame filters, in detail.
For a better comprehension, we arrange all the five of them 
in an increasing order of design complexity. 

In the multi-frame scenario, there exist slightly different types 
of data in the temporal and spatial dimensions. Thus, in order to 
combine them carefully the first two extensions break down 
spatio-temporal filtering into two separable stages. \\\\
\noindent\textit{Proposed Extension - \textbf{A}verage 
then \textbf{F}ilter (AF):} 
First, we average all the frames registered using optical flow. 
Then we employ a single-frame filter for removing the remaining 
noise in the averaged frame. \\\\
\textit{Proposed Extension - 
\textbf{F}ilter then \textbf{A}verage (FA):} Here, we 
first denoise every registered frame by using a single-frame filter 
and then average the denoised frames. \\

The above two approaches differ from some previous methods
\cite{Buades2009, buades2010MultiFrame} in the following 
fundamental aspect: 
Irrespective of the quality of registration, we utilize a temporal 
average \textit{and} spatially filter strategy. This is 
different from a temporal average \textit{or} spatially filter 
technique that depends on the registration error. 
While the first two extensions FA and AF perform a separable 
spatio-temporal filtering, the subsequent three employ 
combined filtering ideas. 
The first two among the three techniques 
utilize 2D patches and the final strategy 
considers 3D spatio-temporal ones. Let us discuss them in 
more detail now.  \\\\
\textit{Existing Extension  - \textbf{S}ingle Reference 
Frame \textbf{F}iltering (SF) \cite{buades2005denoising, 
tico2008multi}:} 
Here, a single frame among all available ones is considered as 
the reference frame. One selects reference patches 
from just this frame. For every reference patch, 
a group of similar patches is formed using information from 
all the frames but not just one. \\\\
\textit{Proposed Extension - \textbf{M}ultiple 
Reference Frame \textbf{F}iltering (MF):} 
The fourth extension differs from SF in three 
different aspects. Firstly, in order to make complete use of the 
available information we consider all frames for reference 
patches. Secondly, we perform an aggregation of denoised pixels 
in such a way that after the first main step we have as many denoised 
frames as there are initial ones. This paves the way for the final 
difference: For every reference patch we find similar patches 
from all frames in the second main step also. 
We cannot do this in the second main step using SF because it has 
considered reference patches from just one frame initially.  
We can thus formulate the final denoised image $\bm{u}^{\textrm{\textbf{final}}}$ which is obtained
from a combination of the registered noisy data $\bm{f}$ and the initial 
denoised image $\bm{u}^{\textrm{\textbf{initial}}}$, as
\begin{equation}
\label{equation_1}
\bm{u}^{\textrm{\textbf{final}}}({\bm{x}}) = 
  \frac{\sum\limits_\ell \sum\limits_{P_\ell} 
  w_{P_\ell}^{\textrm{\textbf{wien}}} 
  \sum\limits_{Q \in P(\textrm{P}_\ell)} \chi_Q(\bm{x}) 
  \bm{u}^{\textrm{\textbf{wien}}}_{Q,P_\ell}(\bm{x})}
  {\sum\limits_{\ell} \sum\limits_{P_\ell} 
  w_{P_\ell}^{\textrm{\textbf{wien}}} 
  \sum\limits_{Q \in P(\textrm{P}_\ell)} \chi_Q(\bm{x}) }.
\end{equation}
Here, $\bm{x}$ denotes the 2D position vector.
We represent the set of most similar patches to the reference patch 
$P_\ell$ belonging to frame $\ell$, using $\mathcal P (P_\ell)$. 
For every patch $Q$ in the set $\mathcal P (P_\ell)$, 
we have $\chi_Q(\bm{x}) = 1 $ if $\bm{x} \in Q$ and 0 otherwise.
The symbol $\bm{u}^{\textrm{\textbf{wien}}}_{Q,P_{\ell}}(\bm{x})$  
denotes the estimation of the value at pixel position $\bm{x}$,  
belonging to the patch $Q$. We derive this estimation through Wiener 
filtering (with coefficients $w_{P_\ell}^{\textrm{\textbf{wien}}}$) 
a combination of $\bm{f}$ and $\bm{u}^{\textrm{\textbf{initial}}}$. 
In similar spirit to \eqref{equation_1},
we can formulate the NLB aggregation process:
\begin{equation}
\label{equation_2}
\bm{u}^{\textrm{\textbf{final}}}({\bm{x}}) = 
  \frac{\sum\limits_\ell \sum\limits_{P_\ell}  
  \sum\limits_{Q \in P(\textrm{P}_\ell)} \chi_Q(\bm{x}) 
  \bm{u}^{\textrm{\textbf{bayes}}}_{Q,P_\ell}(\bm{x})}
  {\sum\limits_{\ell} \sum\limits_{P_\ell} 
  \sum\limits_{Q \in P(\textrm{P}_\ell)} \chi_Q(\bm{x}) }.
\end{equation}
Here, the superscript \textbf{bayes} implies Bayesian 
filtering \cite{LBM2013, LBM2013a}. 
By restricting the total number of frames to one in 
\eqref{equation_1} and \eqref{equation_2}, we obtain the 
original single-frame BM3D and NLB algorithms. This implies that MF 
encompasses the single-frame filters. \\ 

While grouping and filtering stages produce noise-free patches, 
aggregation computes the final denoised image from them. 
Employing 3D spatio-temporal patches gives an advantage 
of having more information at the patch denoising steps itself, 
even before employing the aggregation process. 
This exact idea is employed by the final extension. \\\\
\textit{Existing Extension - \textbf{C}ombined \textbf{F}iltering (CF)
\cite{maggioni2012video, maggioni2012nonlocal, arias2018video}:}  
One fixes 3D spatio-temporal patches and searches for similar volumes
instead of patches. Then, a 4D filtering technique is employed, which 
removes noise using all the considered similar volumes.
Such ideas are in accordance with the single-frame 
NLB and BM3D filters, where one considers a 2D similarity 
measure combined with a 3D denoising technique.\\ 

Table \ref{table8} serves as a look up table for the above five 
extensions and presents the chief characteristics of each one of them.
By combining the five multi-frame extensions 
and the two single-frame filters, we have ten filters in total.
As an example, we will abbreviate one of these combined 
techniques as BM3D-MF, if it is a combination of 
single-frame BM3D with extension MF.
Due to space constraints, within the experimental results that are 
going to be presented in the upcoming subsections, we sometimes 
use shortforms for NLB-MF as NL-MF and BM3D-MF as BM-MF. 
{Moreover, we 
use the abbreviation TA 
to denote temporal averaging. For non-registered data, TA denotes 
averaging after optical flow-based registration.}

%--------------------------------------------------------------% 
\begin{table}[t]
\small
\centering
\setlength{\tabcolsep}{3pt}
\begin{tabular}{ l c }
 \hline 
  Method & Characteristics \\
 %& \% \\
 \hline  
AF & \makecell[l]{1. separable spatio-temporal filtering \\
			   2. \textbf{a}verage 
			   registered frames and then \textbf{f}ilter}\\ 
 \hline			   
FA & \makecell[l]{1. separable spatio-temporal filtering \\
			   2. \textbf{f}ilter 
			   each registered frame and then 
			   				\textbf{a}verage} \\
 \hline			   				
SF & \makecell[l]{1. combined spatio-temporal filtering \\
			   2. considers 2D reference patches from 
			   a \textbf{s}ingle frame} \\
 \hline			   				
MF & \makecell[l]{1. combined spatio-temporal filtering \\
			   2. considers 2D reference patches 
			   from \textbf{m}ultiple frames} \\
 \hline			   
CF & \makecell[l]{1. \textbf{c}ombined spatio-temporal 
			   \textbf{f}iltering \\
			   2. considers 3D reference patches across 
			   frames} \\
                          
   \hline 
 \end{tabular}
 \vspace{2mm}
 \captionof{table}{ The main characteristics 
 of the multi-frame extensions.}
 \label{table8}
\end{table}
%--------------------------------------------------------------%
%--------------------------------------------------------------% 
\begin{table}[t]
\small
\centering
\setlength{\tabcolsep}{3pt}
\begin{tabular}{ l }
% \hline 
%	\textit{\textbf{Algorithm}}: 
%	The proposed image denoising framework \\
 %& \% \\
 \hline  
\textbf{\textit{Input:}} \hspace{1mm} Noisy non-registered 
dataset $\bm{f^{\textrm{nr}}}$ \\ 
\textit{\textbf{Main Algorithm:}} \\
1. We employ an optical flow technique for obtaining
       		 registered data \\ 
\hspace{1.5mm}   $\bm{f}$ from $\bm{f^{\textrm{nr}}}$.
       		 Options for the optical flow methods include \\ 
\hspace{1.5mm} 	 SOF-1, SOF-2 or SOF-3. \\
2. We utilize a combination of single-frame denoising 
filters with their \\ 
\hspace{1.5mm} multi-frame extensions for producing 
       	     the final denoised output $\bm{u}^{\textrm{final}}$ \\
\hspace{1.5mm} using registered data $\bm{f}$. 
Options for the single-frame filters are NLB \\ 
\hspace{1.5mm} or BM3D. They can be 
 combined with extensions AF, FA, SF or MF. \\
\textbf{\textit{Output:}} \hspace{1mm} Denoised data
                          $\bm{u^{\textrm{final}}}$ \\                         
                          
   \hline 
 \end{tabular}
 \vspace{2mm}
 \captionof{table}{A general algorithm of the proposed  
 denoising scheme.}
 \label{table7}
\end{table}
%--------------------------------------------------------------%
%--------------------------------------------------------------%
\begin{figure*}[t]
  \centering
    \includegraphics[width=\linewidth]
  {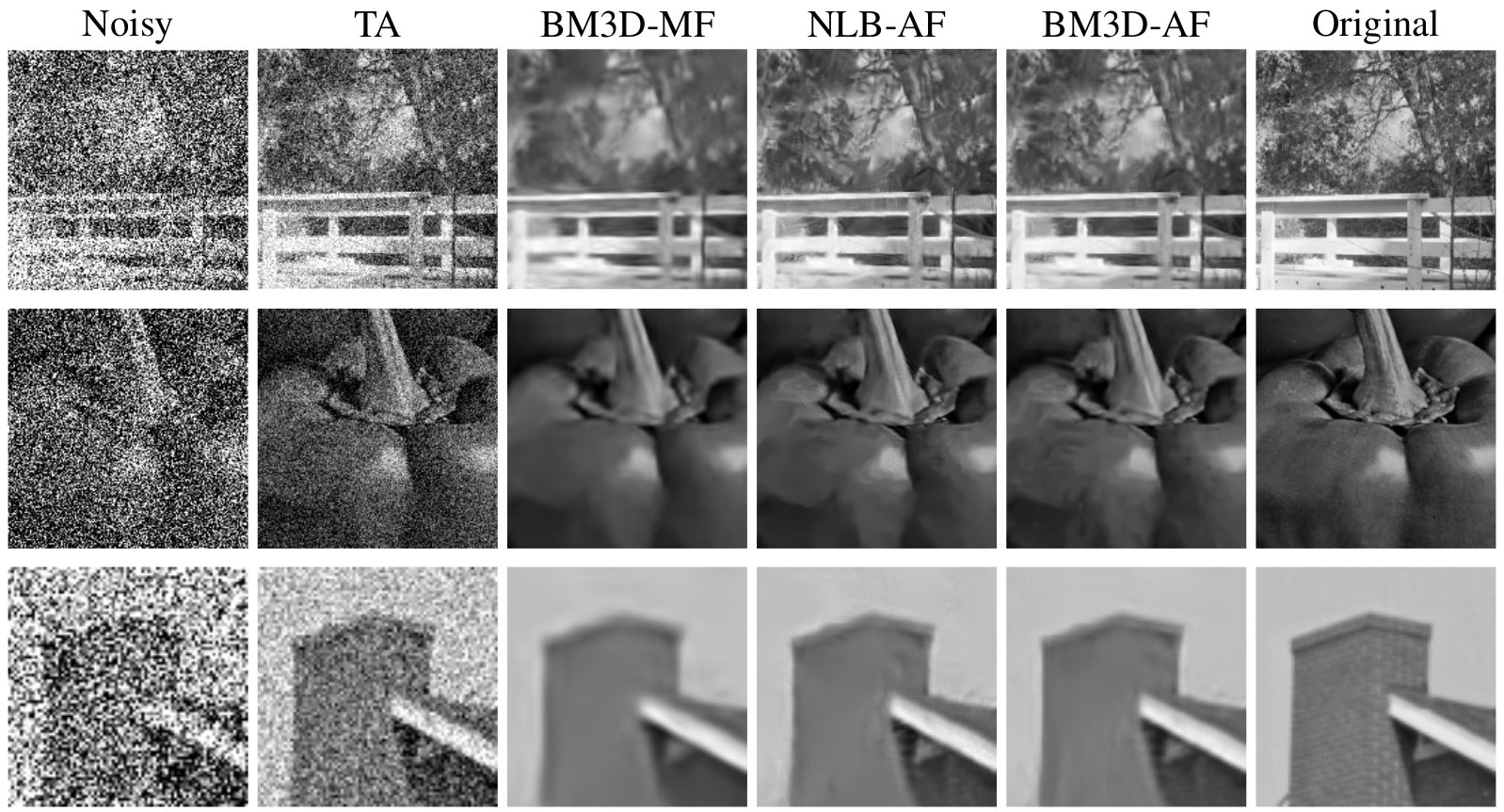}   
\caption{Denoised ten-frame datasets using the three best filters
($\sigma_{\textrm{noise}} = 120$). 
\textbf{Top to Bottom:} Zoom into the Bridge, Peppers 
and House images, respectively.  }
\label{fig:res3}
\end{figure*}
%--------------------------------------------------------------%
\begin{table*}
\small \centering
\setlength{\tabcolsep}{3pt}
\begin{minipage}{\textwidth}
\centering
\begin{tabular}{ l r  r r  r r r r r r r}
 \hline 
 
 Data & NL-AF & NL-FA & NL-SF & NL-MF & NL-CF & 
 BM-AF & BM-FA & BM-SF & BM-MF & BM-CF \\
 %& \% \\
 \hline 
 B10 & \textbf{36.54} & 35.00 & 33.22 & 36.11 & 35.75 & 
 36.53 & 34.16 & 32.54 & 34.91 & 35.95 \\
 B20 & 31.83 & 29.79 & 28.31 & 30.55 & 31.01 & 
 \textbf{31.88} & 28.94 & 28.20 & 29.79 & 31.04 \\  
 B40 & \textbf{27.92} & 25.87 & 24.84 & 26.17 & 26.98 &  
 27.90 & 25.65 & 24.89 & 26.21 & 26.50 \\
 B60 & \textbf{26.04} & 23.98 & 23.11 & 23.86 & 25.07 & 
 25.99 & 24.27 & 23.50 & 24.73 & 24.50 \\
 B80 & \textbf{24.87} & 22.86 & 22.24 & 22.82 & 23.69 & 
 24.83 & 23.51 & 22.75 & 23.88 & 23.45 \\
 B100 & 24.00 & 22.34 & 21.76  & 22.35 & 22.87 & 
 \textbf{24.08} & 22.93 & 22.17 & 23.25 & 22.75 \\  
               \vspace{0.5em}
 B120 & 23.30 & 21.99 & 21.29 & 21.98 & 22.24 & 
 \textbf{23.48} & 22.49 & 21.69 & 22.75 & 22.20 \\ 
 P10 & 38.64 & 37.28 & 36.04 & 37.23 & 37.81 & 
 \textbf{38.72} & 36.90 & 36.04 & 37.11 & 37.47 \\
 P20 & 35.80 & 35.02 & 33.77 & 35.09 & 35.29 & 
 \textbf{35.88} & 34.87 & 33.97 & 35.24 & 34.81\\  
 P40 & 33.23 & 32.49 & 31.10 & 32.66 & 33.00 & 
 \textbf{33.54} & 32.61 & 31.43 & 33.09 & 32.16 \\
 P60 & 31.99 & 30.75 & 29.19 & 30.88 & 31.28 & 
 \textbf{32.10} & 31.13 & 29.79 & 31.63 & 30.41 \\  
 P80 & 30.71 &  29.25 & 28.00 & 29.51 & 29.33 & 
 \textbf{30.84} & 30.34 & 28.53 & 30.41 & 29.12 \\ 
 P100 & 29.76 & 28.32 & 26.97 & 28.65 & 28.28 & 
 \textbf{29.86} & 29.00 & 27.52 & 29.43 & 28.07 \\ 
  \vspace{0.5em}
 P120 & 28.84 & 27.60 & 26.04 & 27.78 & 27.37 & 
 \textbf{28.99} & 28.14 & 26.69 & 28.61 & 27.16\\  
% L10 & 38.84 & 37.80 & 36.38 & 37.89 & 37.96 & 
% \textbf{38.87} & 37.54 & 36.41 & 37.83 & 37.87 \\
% L20 & 35.87 & 34.91 & 33.43 & 34.93 & 35.38 & 
% \textbf{36.00} & 34.75 & 33.75 & 35.27 & 34.83 \\  
% L40 & 32.95 & 31.85 & 30.44 & 32.07 & 32.67 & 
% \textbf{33.23} & 31.90 & 30.71 & 32.53 & 31.60 \\
% L60 & 31.38 & 30.00 & 28.47 & 30.03 & 30.71 & 
% \textbf{31.56} & 30.30 & 29.06 & 30.93 & 29.65 \\ 
% L80 & 30.11 & 28.41 & 27.22 & 28.35 & 28.78 & 
% \textbf{30.15} & 29.04 & 27.83 & 29.59 & 28.28 \\ 
% L100 & 29.13 & 27.34 & 26.17 & 27.50 & 27.75 & 
% \textbf{29.24} & 28.14 & 26.81 & 28.66 & 27.21\\  
%  \vspace{0.5em}
% L120 & 28.24 & 26.77 & 25.41 & 26.96 & 26.87 &  
% \textbf{28.33} & 27.38 & 26.08 & 27.82 & 26.33\\   
 H10 & 39.92 & 38.13 & 36.60 & 37.78 & 39.28 &   
 \textbf{40.12} & 38.15 & 37.23 & 38.75 & 38.79\\
 H20 & 36.36 & 35.20 & 34.02 & 35.32 & 36.33 & 
 \textbf{36.83} & 35.30 & 34.45 & 35.83 & 35.17 \\  
 H40 & 33.22 & 32.58 & 31.22 & 33.23 & 33.46 & 
 \textbf{33.92} & 32.77 & 31.64 & 33.42 & 32.06 \\
 H60 & 31.97 & 30.37 & 28.81 & 31.30 & 31.51 & 
 \textbf{32.49} & 30.96 & 29.77 & 31.77 & 29.94 \\ 
 H80 & 30.52 & 28.26 & 27.23 & 29.20 & 29.49 & 
 \textbf{30.96} & 29.41 & 28.30 & 30.16 & 28.38\\
 H100 & 29.38 & 26.79 & 26.03 & 27.77 & 28.38 & 
 \textbf{29.85} & 28.43 & 27.20 & 29.07 & 27.14\\  
 H120 & 28.46 & 25.66 & 25.08 & 26.71 & 27.35 & 
 \textbf{29.16} & 27.47 & 26.29 & 28.28   & 26.10\\   
 \hline 
 \end{tabular}
\end{minipage}
\vspace{2mm}
\captionof{table}{ PSNR values after denoising five-frame 
 datasets with various 
 methods. Abbreviations: B80 - Bridge with 
 $\sigma_{\mathrm{noise}} = 80$, 
 P - Peppers, H - Bridge. 
 Sizes: H - 256$\times$256, rest - 512$\times$512. }
 \label{table2}
\end{table*}
\begin{table*}
\small \centering
\setlength{\tabcolsep}{3pt}
\begin{minipage}{\textwidth}
\centering
 \begin{tabular}{l r r  r r  r r r r r r r}
 \hline 
 Data & NL-AF & NL-FA & NL-SF & NL-MF & NL-CF & 
 BM-AF & BM-FA & BM-SF & BM-MF & BM-CF \\
 %& \% \\
 \hline 
  B10 & \textbf{39.08} & 35.84 & 33.6700 & 38.40 & 38.41 & 
  39.06 & 34.73 & 33.05 & 37.34 & 37.71 \\  
  B20 & 34.11 & 30.19 & 28.51 & 32.13 & 33.32 & 
  \textbf{34.13} & 29.19 & 28.76 & 31.72 & 32.27 \\  
  B40 & \textbf{29.83} & 26.11 & 24.90 & 26.95 & 28.55 & 
  29.80 & 25.83 & 25.03 & 27.15 & 26.96 \\  
  B60 & \textbf{27.63} & 24.14 & 23.13 & 24.07 & 26.07 & 
  27.61 & 24.45 & 23.58 & 25.46 & 24.75 \\  
  B80 & \textbf{26.31} & 23.00 & 22.31 & 23.01 & 24.38 & 
  26.24 & 23.69 & 22.81 & 24.49 & 23.69 \\  
  B100 & \textbf{25.42} & 22.50 & 21.82 & 22.54 & 23.37 & 
  25.34 & 23.14 & 22.26 & 23.85 & 23.02 \\ 
  \vspace{0.5em}
  B120 & \textbf{24.64} & 22.17 & 21.39 & 22.18 & 22.60 & 
  24.63 & 22.70 & 21.79 & 23.29 & 22.51 \\   
  P10 & 40.50 & 37.58 & 36.14 & 37.71 & 39.16 & 
  \textbf{40.64} & 37.12 & 36.26 & 37.96 & 38.03\\
  P20 & 37.12 & 35.36 & 33.89 & 35.61 & 36.13 & 
  \textbf{37.16} & 35.14 & 34.17 & 35.94 & 35.20\\ 
  P40 & 34.69 & 32.87 & 31.23 & 33.25 & 33.75 & 
  \textbf{34.72} & 33.00 & 31.57 & 33.93 & 32.59 \\
  P60 & 33.06 & 31.14 & 29.29 & 31.48 & 32.03 & 
  \textbf{33.40} & 31.55 & 29.93 & 32.52 & 30.87\\ 
  P80 & 32.12 & 29.63 & 28.18 & 30.02 & 30.18 & 
  \textbf{32.26} & 30.47 & 28.70 & 31.36 & 29.62 \\ 
  P100 & 31.32 & 28.83 & 27.18 & 29.23 & 29.10 & 
  \textbf{31.38} & 29.64 & 27.74 & 30.46 & 28.60 \\  
  \vspace{0.5em}
  P120 & 30.41 & 28.19 & 26.30 & 28.55 & 28.13 & 
  \textbf{30.52} & 28.82 & 27.01 & 29.69 & 27.76 \\ 
%  L10 & 40.53 & 38.18 & 36.50 & 38.62 & 39.26 & 
%  \textbf{40.59} & 37.86 & 36.61 & 38.81 & 38.59 \\
%  L20 & 37.32 & 35.30 & 33.55 & 35.54 & 36.45 & 
%  \textbf{37.38} & 35.09 & 33.96 & 36.27 & 35.42\\ 
%  L40 & 34.54 & 32.21 & 30.52 & 32.71 & 33.70 & 
%  \textbf{34.64} & 32.26 & 30.85 & 33.52 & 32.12 \\
%  L60 & 32.66 & 30.39 & 28.52 & 30.52 & 31.70 & 
%  \textbf{32.99} & 30.68 & 29.20 & 31.87 & 30.20 \\   
%  L80 & 31.62 & 28.73 & 27.35 & 28.76 & 29.73 & 
%  \textbf{31.72} & 29.53 & 28.01 & 30.62 & 28.90 \\ 
%  L100 & 30.69 & 27.79 & 26.38 & 27.98 & 28.59 & 
%  \textbf{30.72} & 28.65 & 27.13 & 29.63 & 27.89 \\ 
%  \vspace{0.5em}
%  L120 & 29.81 & 27.26 & 25.68 & 27.53 & 27.64 & 
%  \textbf{29.87} & 27.95 & 26.35 & 28.84 & 27.05 \\   
  H10 & 41.72 & 38.41 & 36.70 & 38.12 & 40.59 & 
  \textbf{41.89} & 38.38 & 37.42 & 39.82 & 39.53 \\
  H20 & 38.17 & 35.53 & 34.12 & 35.79 & 37.32 & 
  \textbf{38.48} & 35.56 & 34.61 & 36.66 & 35.63  \\ 
  H40 & 34.96 & 33.02 & 31.32 & 34.03 & 34.32 & 
  \textbf{35.41} & 33.16 & 31.91 & 34.37 & 32.57 \\
  H60 & 33.14 & 30.86 & 29.00 & 32.18 & 32.60 & 
  \textbf{33.87} & 31.45 & 29.99 & 32.77 & 30.48  \\
  H80 & 32.14 & 28.79 & 27.65 & 30.08 & 30.61 & 
  \textbf{32.62} & 30.02 & 28.64 & 31.31 & 28.89  \\ 
  H100 & 31.20 & 27.28 & 26.28 & 28.74 & 29.39 & 
  \textbf{31.57} & 29.05 & 27.55 & 30.14 & 27.75  \\  
  H120 & 30.35 & 26.19 & 25.32 & 27.99 & 28.32 & 
  \textbf{30.85} & 28.17 & 26.60 & 29.37 & 26.76  \\    
 \hline
\end{tabular}
\end{minipage}
 \vspace{2mm}
 \captionof{table}{
 PSNR values after denoising ten-frame datasets with various 
 methods. Abbreviations as in Table \ref{table2}.}
 \label{table2__1}
\end{table*}
%--------------------------------------------------------------% 
%--------------------------------------------------------------% 
\begin{table*}[t]
\small
\setlength{\tabcolsep}{3pt}
\centering
%\begin{minipage}{0.2\textwidth}
%\begin{tabular}{ l c  c c c}
% \hline 
% 
% Image ($\sigma_\mathrm{noise}$) & $\alpha$ & $\gamma$ & $\lambda$ 
% & AL2E \\
% %& \% \\
% \hline 
% Grove2 (10) & 15 & 1.5 & 0.1 & 0.263 \\  
% Grove2 (20) & 25 & 1.5 & 0.1 & 0.403 \\  
% Grove2 (40) & 35 & 1.5 & 0.1 & 0.520 \\  
% Grove2 (60) & 35 & 1.5 & 0.1 & 0.497 \\  
% Grove2 (80) & 45 &  2.5 & 0.1 & 0.612 \\  
% Grove2 (100) & 35 & 2.0 & 0.1 & 0.669 \\  
% Grove2 (120) & 20 & 1.0 & 0.1 & 0.742 \\   
% \hline 
% \end{tabular}
%\end{minipage}
%\hspace{4em}
%\vspace{1em}
%\begin{minipage}{0.11\textwidth}
\begin{minipage}{0.2\textwidth}
\begin{tabular}{ l c  c c c}
 \hline 
 
 Data & $\alpha$ & $\gamma$ & $\lambda$ 
 & Best Method \\
 %& \% \\
 \hline 
 G10 & 15 & 1.5 & 0.1 & SOF-2 \\  
 G20 & 25 & 1.5 & 0.1 & SOF-2 \\  
 G40 & 35 & 1.5 & 0.1 & SOF-2 \\  
 G60 & 35 & 1.5 & 0.1 & SOF-2 \\  
 G80 & 45 &  2.5 & 0.1 & SOF-2 \\  
 G100 & 110 & 1.0 &  - & SOF-3 \\  
 G120 & 95 & 1.0 &  - & SOF-3 \\   
 \hline 
 \end{tabular}
\end{minipage}
\hspace{8em}
\vspace{1em}
%\begin{minipage}{0.11\textwidth}
% \begin{tabular}{c  c c  c }
% \hline 
% $\alpha$ & $\gamma$ & $\lambda$ & AL2E \\
% %& \% \\
% \hline 
%  15 & 1.5 & 0.1 & \textbf{0.263} \\
%  25 & 1.5 & 0.1 & \textbf{0.403} \\
%  35 & 1.5 & 0.1 & \textbf{0.520} \\
%  35 & 1.5 & 0.1 & \textbf{0.497} \\
%  75 & 4.5 & 0.1 & \textbf{0.612} \\  
%  15 & 0.5 & 0.1 & 0.667 \\ 
%  20 & 1.0 & 0.1 & 0.742 \\   
% \hline
%\end{tabular}
%\end{minipage}
%\hspace{3em}
%\vspace{1em}
%\begin{minipage}{0.13\textwidth}
% \begin{tabular}{c  c c  c }
% \hline 
% $\alpha$ & $\gamma$ & AL2E \\
% %& \% \\
% \hline
%  20 & 1.0 & 0.292 \\  
%  75 & 1.0 & 0.410 \\  
%  120 & 1.0 & 0.531 \\  
%  120 & 1.0 & 0.533 \\
%  125 & 1.0 & 0.640 \\  
%  110 & 1.0 & \textbf{0.632} \\ 
%  95 & 1.0 & \textbf{0.711} \\   
% \hline
%\end{tabular}
%\end{minipage}
\begin{minipage}{0.18\textwidth}
 \begin{tabular}{l c  c }
 \hline 
 Data & $\alpha$ & $\gamma$ \\
 %& \% \\
 \hline 
  S10 & 25 & 1.5 \\  
  S20 & 75 & 2.5 \\  
  S40 & 95 & 1.5 \\ 
  S60 & 110 & 0.5 \\  
  S80 &  85 & 0.5 \\  
  S100 &  95 & 0.5 \\ 
  S120 &  90 & 0.5 \\   
 \hline
\end{tabular}
\end{minipage}
\hspace{0.5em}
\begin{minipage}{0.18\textwidth}
 \begin{tabular}{l c  c }
 \hline 
 Data & $\alpha$ & $\gamma$ \\
 %& \% \\
 \hline 
  BH10 &  100 & 0.5 \\  
  BH20 &  130 & 0.5 \\  
  BH40 &  135 & 1.0 \\ 
  BH60 &  135 & 0.5 \\   
  BH80 &  130 & 1.5 \\  
  BH100 & 100 & 1.5 \\ 
  BH120 &   90 & 1.5 \\   
 \hline
\end{tabular}
\end{minipage}

 \captionof{table}{Optical flow parameter values used for 
 different datasets. 
 \textbf{Left:} Grove2 
 dataset with the best among SOF-1, SOF-2 and SOF-3 methods. 
 We have 
 considered the tenth  frame as the reference frame  since ground truth 
 flow information was available between frames 10 and 11.
 \textbf{Centre:} Shoe dataset with SOF-3 approach.  
 \textbf{Right:} Bird House dataset with SOF-3 technique. We have 
 utilized the fifth frame as the reference frame for the latter two 
 datasets and then employed frames 4-6 for optimizing the optical 
 flow parameters. Also, we have used BM3D-MF and BM3D-FA as denoising 
 filters for optimizing SOF parameters for these two
 datasets, respectively.}
 \label{table3}
\end{table*}
%--------------------------------------------------------------% 

\subsection{Optical Flow Methods Used}
As already mentioned, we perform experiments on both 
perfectly registered and non-registered datasets. 
In the latter scenario, we need to first register the images 
before applying the above multi-frame extensions.
Thus, we have employed three robust discontinuity preserving optical 
flow methods \cite{Alvarez1999, monzon2016regularization, 
monzon2016IPOL}. These motion estimation techniques 
perform better than some classical strategies 
\cite{BBPW04, zach2007duality}. 
In all the three approaches, one minimizes a similar energy 
functional to determine the motion vector 
$\bm{w} = (w_1,w_2,1)^\top$ between frames 
$f_1$ and $f_2$:
\begin{equation}
\begin{split}
  E(\bm{w}) & =
 \int_{\Omega} \Big( \Psi (|f_2(\bm{x} + \bm{w}) 
 - f_1(\bm{x})|^2) +  \\&
 \gamma \left( \Psi (|\bm{\nabla}f_2(\bm{x} 
 + \bm{w} \right)  -  \bm{\nabla}f_1(\bm{x})|^2) + \\& 
 \alpha \left( \Psi (\Phi\left(\bm{\nabla}
 f_1(\bm{x})\right) 
 \cdot \left(|\bm{\nabla} w_1|^2 + 
 |\bm{\nabla} w_2|^2\right) \right)\Big)\ d\bm{x}.
\end{split} 
\end{equation}
Here, $\bm{x} = (x, y, t)^T$ denotes the spatio-temporal location, 
$\Omega$ is the 2D image domain and $\bm{\nabla}$ is the 
spatio-temporal gradient. The above energy penalizes
deviations in both gray values and gradients. 
One enables interactions in between neighboring pixels 
through the smoothness term. 
The parameters $\gamma$ and $\alpha$ represent the 
gradient and smoothness term weights, respectively.
Moreover, applying
$\Psi (s^2) = \sqrt{s^2 + \epsilon^2}$ results in a robust 
convex energy functional with $\epsilon = 0.001$ ensuring strict 
convexity of $\Psi$. 
The smoothness function $\Phi(\bm{\nabla}f_1,\lambda)$ 
with parameter $\lambda$ specifies the regularisation strategy. 
The three optical flow methods 
that we use in this work differ in the 
choice of this particular function.
We abbreviate these three techniques as SOF-1, -2 
and -3 (SOF means sub-optimal flow). In SOF-1, one employs a 
decreasing scalar function $\Phi(\bm{\nabla}f_1,\lambda)$ 
to preserve image driven flow discontinuities. 
The second and third optical flow strategies 
try to avoid blob like artifacts using two different approaches. 
SOF-2 performs a minimum 
isotropic diffusion even when the gradient is very large. 
In SOF-3, one utilizes an automatic selection strategy 
for $\lambda$. The same numerical procedure is adopted 
to compute the solution in all the three methods.
%--------------------------------------------------------------%
\begin{table*}[t]
\small
\centering
\setlength{\tabcolsep}{3pt}
\begin{tabular}{ l r  r r  r r r r r r r r r}
 \hline 
	Data & NL-AF & NL-FA & NL-SF 
	& NL-MF & NL-CF & BM-AF & BM-FA & BM-SF & BM-MF & BM-CF \\
 %& \% \\
 \hline 
 G10 & 33.10 & 31.80 & 32.23 & 32.16 & \textbf{34.14} &
 32.89 & 31.50 & 31.93 & 31.80 & 33.22 \\  
 G20 & 30.24 & 28.62 & 28.27 & 28.75 & \textbf{30.58} & 
 30.09 & 28.20 & 28.14 & 28.64 & 29.74 \\  
 G40 & \textbf{27.26} & 25.02 & 24.37 & 24.72 & 27.06 & 
 27.03 & 25.52 & 25.12 & 25.82 & 26.15 \\  
 G60 & \textbf{25.32} & 23.75 & 23.20 & 23.60 & 25.22 & 
 25.32 & 24.36 & 23.82 & 24.54 & 24.42 \\ 
 G80 & 24.05 & 23.07 & 22.66 & 23.06 & 23.93 & 
 \textbf{24.39} & 23.68 & 23.17 & 23.85 & 23.45 \\  
 G100 & 23.21 & 22.65 & 22.15 & 22.65 & 23.17 & 
 \textbf{23.60} & 23.13 & 22.52 & 23.27 & 22.79 \\  
               \vspace{0.5em}
 G120 & 22.76 & 22.41 & 21.81 & 22.41 & 22.58 & 
 \textbf{23.10} & 22.76 & 22.15 & 22.87 & 22.28\\ 
  \hline 
 G10  & 33.21 & 31.39 & 32.41 & 32.40 & \textbf{35.46} & 
 33.04 & 31.11 & 32.22 & 32.17 & 33.29 \\  
 G20  & 30.83 & 28.44 & 28.40 & 29.27 & \textbf{31.90} & 
 30.74 & 28.04 & 28.60 & 29.45 & 29.88 \\  
 G40  & 27.97 & 24.85 & 24.39 & 24.77 & \textbf{28.23} & 
 27.83 & 25.46 & 25.24 & 26.31 & 26.22 \\  
 G60  & \textbf{26.18} & 23.61 & 23.22 & 23.57 & 26.10 & 
 26.04 & 24.34 & 23.87 & 24.93 & 24.48 \\ 
 G80  & 24.97 & 23.01 & 22.64 & 23.07 & 24.61 & 
 \textbf{24.97} & 23.70 & 23.20 & 24.13 & 23.51 \\  
 G100 & 23.99 & 22.73 & 22.19 & 22.76 & 23.71 & 
 \textbf{24.12} & 23.24 & 22.56 & 23.57 & 22.88\\   
  \vspace{0.5em} 
 G120 & 23.25 & 22.51 & 21.87 & 22.51 & 22.76 & 
 \textbf{23.48} & 22.89 & 22.18 & 23.12 & 22.40\\   
  \hline 
 \end{tabular}
 \vspace{1em}
 \captionof{table}{PSNR values of denoised Grove2 images after using a 
 combination of denoising methods and optical flow. \textbf{Top:} 
 Four-frame datasets (frames 9-12). \textbf{Bottom:} Eight-frame 
 datasets (frames 7-14). Frame size: 640 $\times$ 480.}
 \label{table4}
\end{table*}
%--------------------------------------------------------------%
%--------------------------------------------------------------%
\begin{table*}[t]
\small
\centering
\setlength{\tabcolsep}{3pt}
\begin{tabular}{ l r  r r  r r r r r r r r r}
 \hline 
 Data & NL-AF & NL-FA & NL-SF & NL-MF  
 & NL-CF & BM-AF & BM-FA & BM-SF & BM-MF & BM-MFO & BM-CF\\
 %& \% \\
 \hline 
 S10 & 37.49 & 36.34 & 35.94 & 36.38 & \textbf{37.89} & 
 37.67 & 36.84 & 36.51 & 36.98 & 36.79 & 37.38 \\  
 S20 & 34.63 & 33.32 & 32.63 & 33.32 & 35.02 & 
 \textbf{35.02} & 34.10 & 33.39 & 34.35 & 34.16 & 34.28 \\  
 S40 & 31.71 & 30.17 & 29.51 & 30.27 & 32.08 & 
 \textbf{32.20} & 31.37 & 30.46 & 31.74 & 31.63 & 31.16 \\  
 S60 & 30.39 & 28.58 & 27.77 & 28.77 & 30.26 & 
 \textbf{30.90} & 29.84 & 28.80 & 30.25 & 30.18 & 29.34  \\ 
 S80 & 29.07 & 27.46 & 26.64 & 27.66 & 28.52 & 
 \textbf{29.65} & 28.71 & 27.59 & 29.05 & 29.06 & 28.01 \\  
 S100 & 28.27 & 26.80 & 25.88 & 27.08 & 27.52 & 
 \textbf{28.88} & 27.88 & 26.65 & 28.09 & 28.26 & 26.95 \\  
               \vspace{0.5em}
 S120 & 27.61 & 26.35 & 25.26 & 26.64 & 26.70 & 
 \textbf{28.14} & 27.18 & 25.90 & 27.35 & 27.51 & 26.08 \\ 
  \hline 
 S10  & 37.55 & 35.95 & 35.90 & 36.28 & \textbf{38.09} & 
 37.66 & 36.48 & 36.56 & 37.03 & 36.83 & 37.42 \\  
 S20  & 35.19 & 33.22 & 32.67 & 33.39 & 35.26 & 
 \textbf{35.45} & 34.02 & 33.51 & 34.74 & 34.58 & 34.44 \\  
 S40  & 32.47 & 30.19 & 29.61 & 30.45 & 32.38 & 
 \textbf{32.87} & 31.56 & 30.63 & 32.12 & 32.29 & 31.49 \\  
 S60  & 31.28 & 28.65 & 27.84 & 29.05 & 30.61 & 
 \textbf{31.79} & 30.10 & 28.97 & 30.93 & 30.97 & 29.77 \\ 
 S80 & 30.13 & 27.56 & 26.75 & 28.02 & 29.01 & 
 \textbf{30.62} & 29.06 & 27.75 & 29.78 & 29.90 & 28.53 \\  
 S100 & 29.34 & 27.01 & 25.98 & 27.49 & 28.02 & 
 \textbf{29.93} & 28.27 & 26.85 & 28.87 & 29.11 & 27.57 \\   
  \vspace{0.5em} 
 S120 & 28.68 & 26.66 & 25.37 & 27.10 & 27.19 & 
 \textbf{29.27} & 27.65 & 26.04 & 28.01 & 28.35 & 26.77 \\   
  \hline 
 \end{tabular}
 \vspace{1em}
 \captionof{table}{PSNR values of denoised Shoe images after using a 
 combination of denoising methods and optical flow. \textbf{Top:} 
 Five-frame datasets (frames 3-7).  \textbf{Bottom:} Ten-frame 
 datasets (frames 1-10). Frame size: 1280 $\times$ 720. 
 \textbf{Abbreviation:} BM-MFO uses twice the number of 
 patches as in BM-MF. }
 \label{table5}
\end{table*}
%--------------------------------------------------------------%
%--------------------------------------------------------------% 
\begin{table*}[t]
\small
\centering
\setlength{\tabcolsep}{3pt}
\begin{tabular}{ l r  r r  r r r r r r r r}
 \hline 
 Data & NL-AF & NL-FA & NL-SF & NL-MF 
 & NL-CF & BM-AF & BM-FA & BM-SF & BM-MF & BM-CF\\
 %& \% \\
 \hline 
 BH10  & 36.63 & 35.00 & 34.86 & 35.61 & 35.03 & 
 \textbf{36.63} & 34.99 & 34.84 & 35.37 & 35.07 \\  
 BH20  & 33.46 & 31.21 & 30.67 & 31.73 & 31.57 & 
 \textbf{33.55} & 31.19 & 30.88 & 31.84 & 31.60  \\  
 BH40  & 30.07 & 27.02 & 26.43 & 27.11 & 28.22 & 
 \textbf{30.11} & 27.84 & 27.36 & 28.48 & 27.79  \\  
 BH60  & 28.15 & 25.16 & 24.52 & 24.87 & 26.52 & 
 \textbf{28.26} & 26.36 & 25.71 & 26.80 & 25.95 \\ 
 BH80 & 26.71 & 24.50 & 24.05 & 24.38 & 25.34 & 
 \textbf{26.95} & 25.53 & 24.88 & 25.89 & 24.90  \\  
 BH100 & 25.73 & 24.18 & 23.73 & 24.13 & 24.62 & 
 \textbf{26.03} & 24.97 & 24.29 & 25.24 & 24.19  \\  
               \vspace{0.5em}
 BH120 & 24.96 & 23.96 & 23.43 & 23.94 & 24.08 & 
 \textbf{25.27} & 24.55 & 23.82 & 24.76 & 23.66  \\ 
  \hline 
 BH10 & \textbf{36.13} & 34.36 & 34.90 & 35.27 & 35.62 & 
 36.12 & 34.32 & 34.94 & 35.23 & 35.19  \\  
 BH20 & 33.79 & 30.94 & 30.74 & 31.97 & 31.97 & 
 \textbf{33.84} & 30.81 & 31.25 & 32.39 & 33.29  \\  
 BH40 & 30.89 & 26.88 & 26.54 & 27.42 & 28.47 & 
 \textbf{31.00} & 27.72 & 27.60 & 29.16 & 27.99  \\  
 BH60 & 29.18 & 25.06 & 24.55 & 24.92 & 26.66 & 
 \textbf{29.24} & 26.37 & 25.79 & 27.33 & 26.11  \\ 
 BH80 & 27.80 & 24.50 & 24.11 & 24.44 & 25.49 & 
 \textbf{27.91} & 25.40 & 24.99 & 26.42 & 25.05  \\  
 BH100 & 26.71 & 24.28 & 23.80 & 24.25 & 24.75 & 
 \textbf{26.84} & 24.70 & 24.39 & 25.76 & 24.39  \\   
  \vspace{0.5em} 
 BH120 & 25.86 & 24.12 & 23.53 & 24.08 & 24.19 & 
 \textbf{25.91} & 24.66 & 23.96 & 25.23 & 23.91  \\   
  \hline 
 \end{tabular}
 \vspace{1em}
 \captionof{table}{PSNR values of denoised Bird House images after 
 using a  combination of denoising methods and optical flow. 
 \textbf{Top:}  Five-frame datasets (frames 3-7). \textbf{Bottom:} 
 Ten-frame datasets (frames 1-10). Frame size: 1280 $\times$ 720.}
 \label{table6}
\end{table*}
%--------------------------------------------------------------%

We use the above mentioned optical flow strategies for the 
first four extensions. The algorithm in Table \ref{table7} 
describes the main ideas behind the denoising framework of 
our approaches. The fifth method CF uses its own motion compensation 
techniques. The difference in the various motion estimation approaches 
used should not be an issue as we are also performing experiments 
on perfectly registered data. This finishes the modeling 
and theory part of this work. Now, we move on to the 
experimental demonstrations.

% -----------------------------------------------------------%
\section{Experiments and Discussion}
\label{sec:exp}
\subsection{Datasets}
For creating perfectly registered data, we have considered multiple 
AWGN realisations of the classical House, Peppers and 
Bridge (http://sipi.usc.edu/database/) images  
with fourteen datasets each. They are obtained by a 
combination of $\sigma_{\textrm{noise}} 
= 10, 20, 40, 60, 80,100,120$ with five- and ten-frame datasets.
In a similar spirit, we have also created non-registered data 
by corrupting the Grove2\cite{baker2011database}, 
Shoe and Bird House \cite{ummenhofer2012dense} images with AWGN.
{ 
It has to be noted that we have not clipped the dynamic range 
of the images after degrading them by noise.}
\subsection{Parameter Selection}
\noindent\textit{Optical Flow Parameters:} For the Grove2 dataset, 
we have optimized the optical flow parameters  
with respect to the ground truth flow for all three 
methods. We then choose the best method to register every dataset. 
For Shoe and Bird House datasets we have optimized the SOF-3 
parameters with respect to the final denoised image directly as 
the ground truth flow was not available. Table \ref{table3} shows more 
details. \newline 
 %--------------------------------------------------------------%
\begin{figure*}[t]
  \centering
    \includegraphics[width=\linewidth]
  {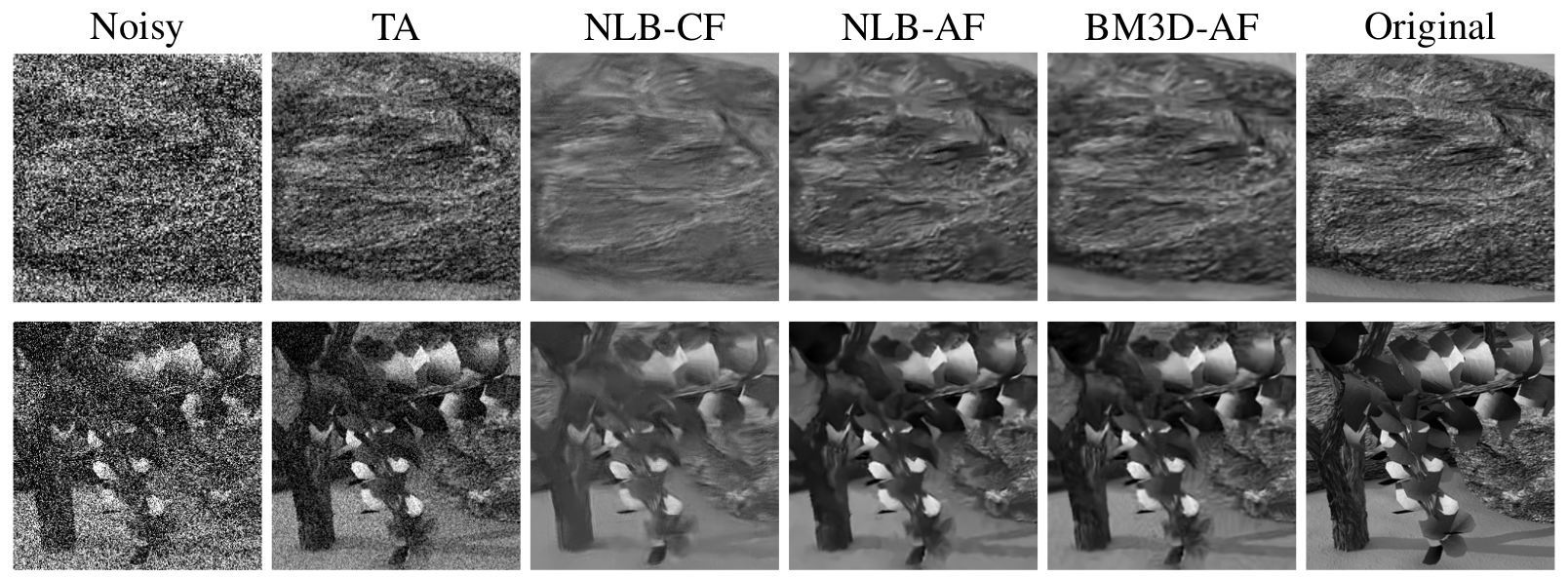}     \\ [0.4em]  
\caption{Denoised regions of 8-frame Grove2 dataset 
using the three best extensions
($\sigma_{\textrm{noise}} = 80$). }
\label{fig:res4}
\end{figure*}
%--------------------------------------------------------------%
%--------------------------------------------------------------% 
\begin{figure*}[t]
  \centering
    \includegraphics[width=\linewidth]
  {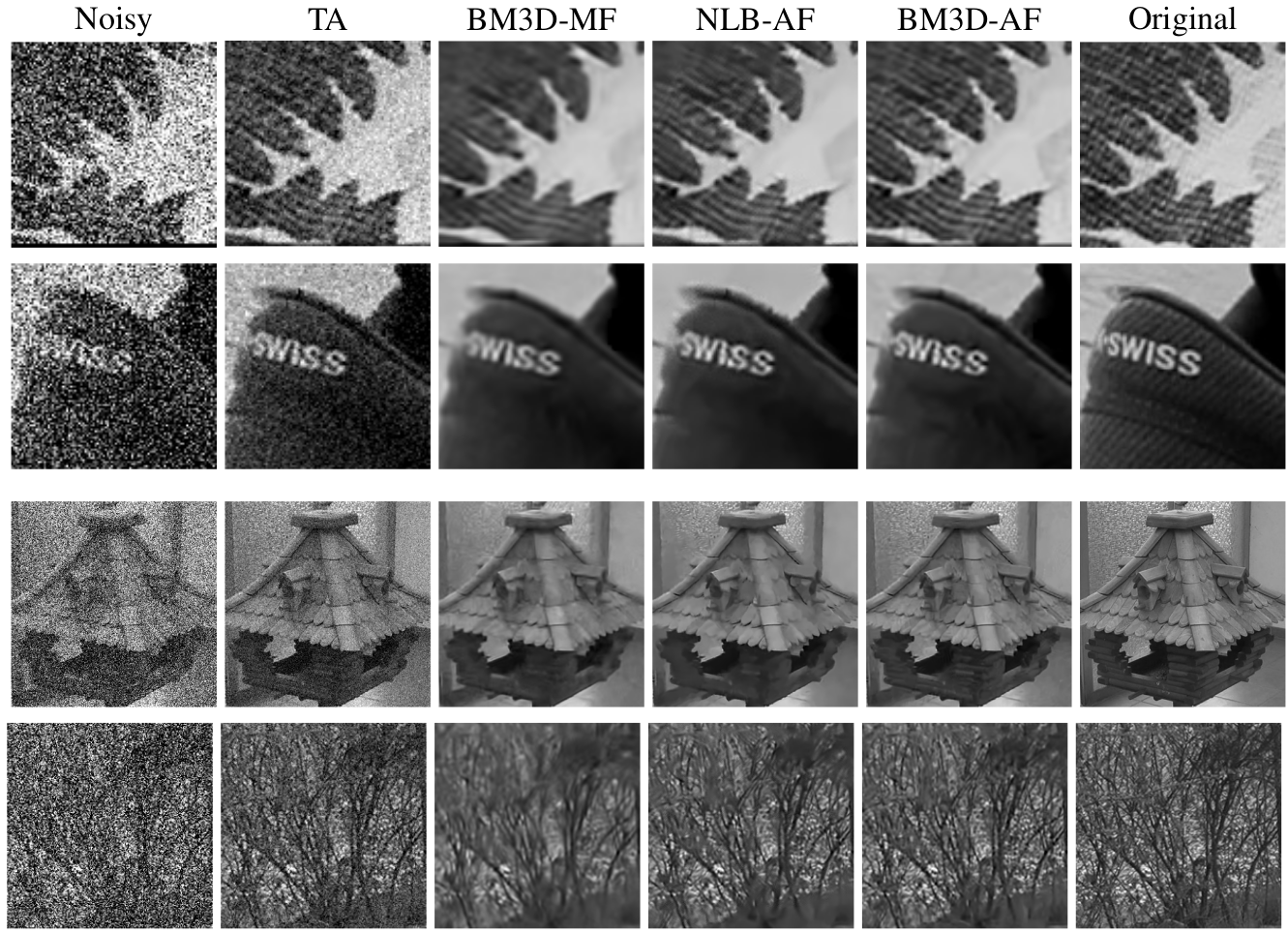}   
  \caption{Different denoised regions of the Shoe (top) and 
Bird House (bottom) datasets using the three best filters
($\sigma_{\textrm{noise}} = 80$). }
\label{fig:res7}
\end{figure*}
%--------------------------------------------------------------% 

\noindent\textit{Denoising Parameters:}
Various studies \cite{LBM2013, LBM2013a, DFKE2007, Lebrun2012, 
hou2010comments} have contributed in making the single-frame filters 
BM3D and NLB parameter selection-free, while retaining the quality of 
the denoised images as much as possible. 
In a similar spirit to the above works, in this paper we 
use better versions of two extensions introduced in 
our conference paper \cite{bodduna2019MultiFrame}. 

Firstly, at the time of application of 
the filter in the first extension AF, the noise distribution has 
already changed due to temporal averaging. Since we are using an 
AWGN model, we know that the standard deviation of noise is reduced 
by a factor $\sqrt{L}$ for a dataset with $L$ frames. 
We can improve the performance of type-AF extensions if we select 
the filter parameters corresponding to the new standard 
deviation. 

The second improvement is to optimize the number 
of patches in a 3D group using 
both the original single-frame BM3D filter as well as 
the BM3D-MF technique. The threshold parameter 
on $L_2$ distance and the parameter which decides the 
maximum patches in a 3D group together 
control the total number of patches one employs for 
filtering purposes. Our experience suggests that the gain in quality  
due to the threshold for low amplitude 
noise elimination, is relatively lot less when compared 
to the deteriotion because of it in case of large noise levels. 
Since one of the main objectives of this paper is to concentrate on 
large noise amplitudes as well, for simplicity reasons 
we refrain from using the threshold 
parameter in any of the first four BM3D extensions. 
Moreover, in the multi-frame scenario we have more similar 
patches, when compared to the single-frame layout. 
We thus check in the upcoming sections, whether 
the best performing extension (BM3D-MF) in our conference 
publication \cite{bodduna2019MultiFrame}, can give even better 
results by increasing the maximum number of patches in a 3D 
group through doubling. We label this particular 
parametric choice as BM3D-MFO, where O stands for an 
optimized version.

For the results of perfectly registered 
noisy data using SF and CF techniques,
we have always presented the best peak signal to noise ratio 
(PSNR) value among all frames.
This ensures a fair comparison with the 
remaining three extensions.
 
For experiments on non-registered datasets, 
we have calculated the PSNR 
value by leaving out a border of 
fifty pixels on all sides of the reference frame at which 
different frames were registered. We do this in order to mitigate 
the ill-effects due to unavailable information at the borders 
of registered images. This also makes sense for several multi-frame 
imaging applications where we capture the region of interest 
in the centre of the frame.
% -----------------------------------------------------------%
%--------------------------------------------------------------%
\begin{figure*}[t]
  \centering
    \includegraphics[width=0.88\linewidth]
  {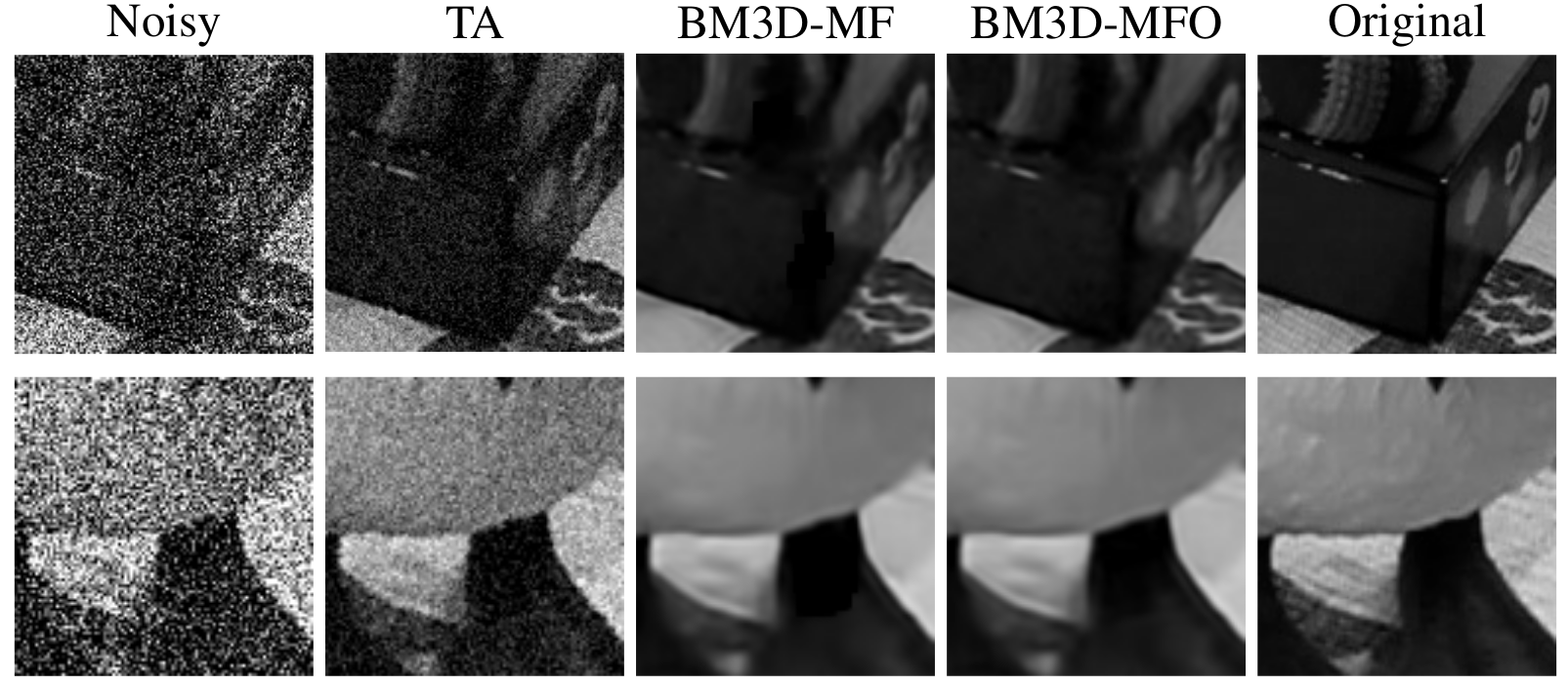}   
\caption{Various denoised regions of the Shoe dataset 
($\sigma_{\textrm{noise}} = 80$). }
\label{fig:res6}
\end{figure*}
%--------------------------------------------------------------%
%--------------------------------------------------------------% 
\begin{table}[t]
\small
\setlength{\tabcolsep}{3pt}
\centering
\begin{tabular}{ l r  r r}
 \hline 
 
 Data & NL-AF & BM-AF & VNLNET\\
 %& \% \\
 \hline 
 B10 &  \textbf{39.08} & 39.06 & 38.35\\
 B20 &  34.11 & \textbf{34.13} & 33.36\\  
               \vspace{0.5em}
 B40 &  \textbf{29.83} & 29.80 & 29.00\\
 P10 &  40.50 & \textbf{40.64} & 38.64\\
 P20 &  37.12 & \textbf{37.16} & 35.88\\  
  			   \vspace{0.5em}
 P40 &  34.69 & \textbf{34.72} & 33.30\\
% L10 &  40.53 & \textbf{40.59} &  39.36\\
% L20 &  37.32 & \textbf{37.38} &  36.38\\  
%  			   \vspace{0.5em}
% L40 &  34.54 & \textbf{34.64} & 33.25\\
 H10 &  41.72 & \textbf{41.89} & 40.50\\
 H20 &  38.17 & \textbf{38.48} & 36.76\\  
 H40 &  34.96 & \textbf{35.41} & 34.01\\
 \hline 
 \end{tabular}
\hspace{7.5mm}
{\begin{tabular}{l r  r r}
 \hline 
 Data & NL-AF & BM-AF & VNLNET \\
 %& \% \\
 \hline 
 G10 &  33.21 & 33.04 & \textbf{34.64}\\
 G20 &  30.83 & 30.74 & \textbf{31.02} \\  
               \vspace{0.5em}
 G40 &  \textbf{27.97} & 27.83 & 27.73\\
 S10 &  37.55 & 37.66 & \textbf{38.39}\\
 S20 &  34.63 & \textbf{35.45} & 35.36\\  
  			   \vspace{0.5em}
 S40 &  31.71 & \textbf{32.87} & 32.37\\
 BH10 & 36.13 & 36.12 & \textbf{37.10}\\
 BH20 & 33.79 & \textbf{33.84} & 33.78\\ 
 BH40 & 30.89 & \textbf{31.00} & 30.14\\
 \hline 
\end{tabular}}
 \vspace{1em}
 \captionof{table}{PSNRs after denoising 10-frame datasets 
 with various methods.  \textbf{Left:} Perfectly registered datasets. 
 \textbf{Right:} Non-registered layout. 
 Abbreviations are as in Table \ref{table2}. 
 Moreover, G stands for Grove2, 
 S for Shoe, and BH for Bird House. }
 \label{table2_1}
\end{table}
%--------------------------------------------------------------% 
%--------------------------------------------------------------% 
\begin{figure*}[t]
  \centering
    \includegraphics[width=0.88\linewidth]
  {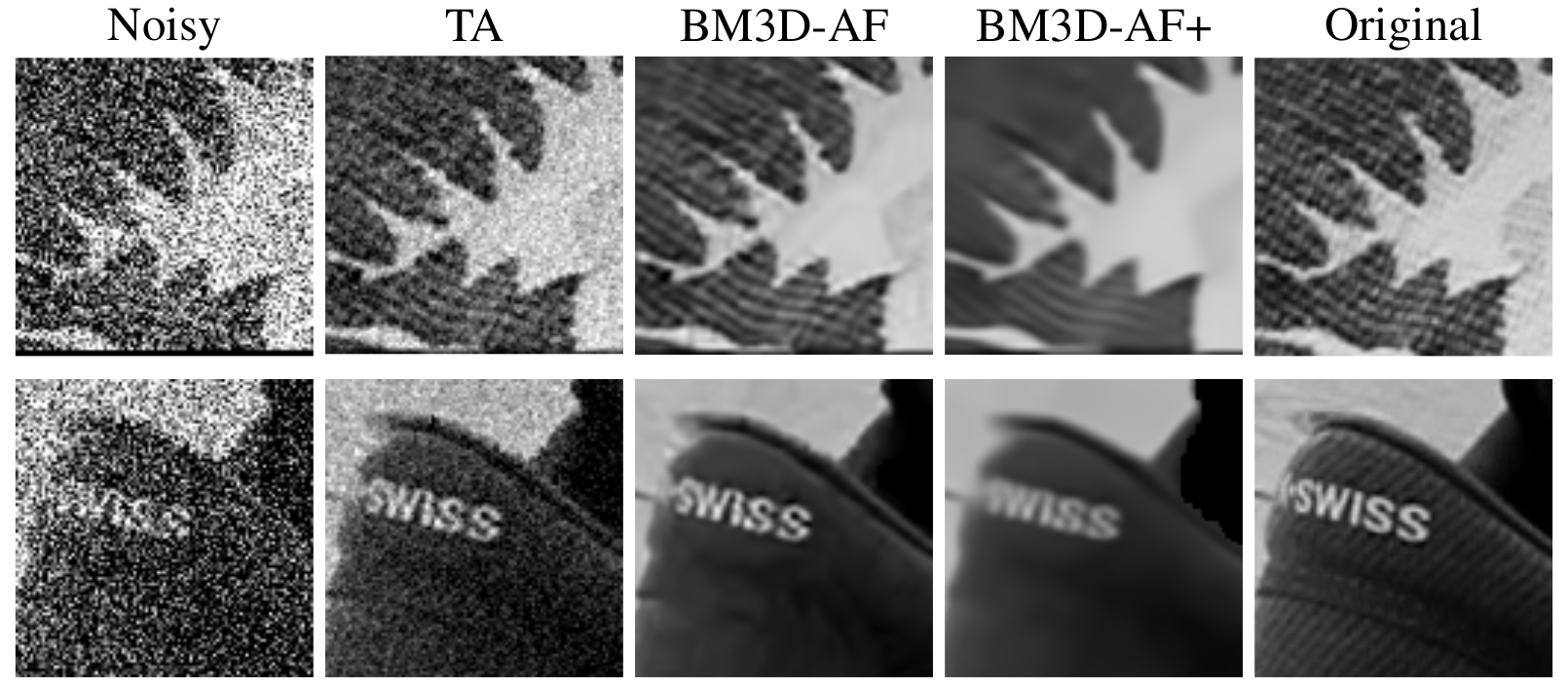}
\vspace{2mm}
\caption{
The BM3D-AF+ variant uses a $\sigma_{\textrm{noise}}$ value that 
corresponds to the raw noisy images. 
It does not consider the change in noise 
distribution due to temporal averaging. 
It produces a result that is inferior by 1.74 dB on a 10-frame dataset 
with $\sigma_{\textrm{noise}} = 80$.}
\label{fig:res7_1}
\end{figure*}
%--------------------------------------------------------------% 
%--------------------------------------------------------------% 
\begin{figure*}[t]
  \centering
    \includegraphics[width=0.88\linewidth]
  {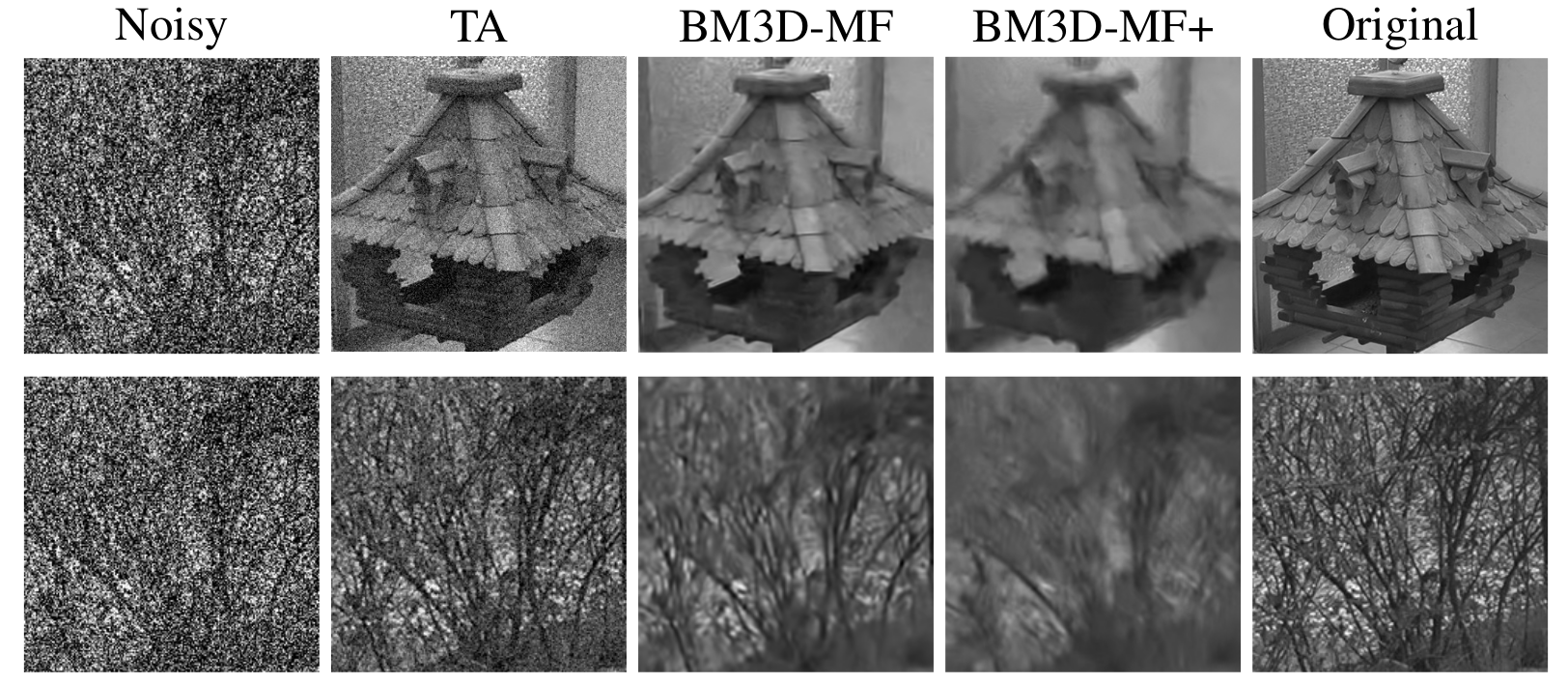}
\vspace{2mm}
\caption{
The BM3D-MF+ variant does not use optical flow-based registration. 
It produces a result that is inferior by 2.36 dB on a 10-frame dataset 
with $\sigma_{\textrm{noise}} = 80$.}
\label{fig:res7_2}
\end{figure*}
%--------------------------------------------------------------%
%--------------------------------------------------------------% 
\begin{table}[t]
\small
\setlength{\tabcolsep}{3pt}
\centering
\begin{tabular}{l r  r r}
 \hline 
 Data & BM-AF & BM-AF+ & Difference \\
 %& \% \\
 \hline 
 G10 & 32.93 & 30.62 & 2.31\\
 G20 & 30.34 & 27.30 & 3.04 \\  
               \vspace{0.5em}
 G40 & 27.33 & 24.97 & 2.36\\
 S10 & 37.67 & 36.22 & 1.45\\
 S20 & 35.02 & 33.51 & 1.51\\  
  			   \vspace{0.5em}
 S40 & 32.20 & 31.01 & 1.19\\
 BH10 & 36.63 & 34.26 & 2.37\\
 BH20 & 33.55 & 30.32 & 3.23\\ 
 BH40 & 30.11 & 27.22 & 2.89\\
 \hline 
\end{tabular}
\hspace{7.5mm}
{\begin{tabular}{l r  r r}
 \hline 
 Data & BM-MF & BM-MF+ & Difference \\
 %& \% \\
 \hline 
 G10 &  31.52 & 22.44 & 9.08\\
 G20 &  28.75 & 22.39 & 6.36\\  
               \vspace{0.5em}
 G40 &  26.01 & 22.23 & 3.78 \\
 S10 &  36.98 & 33.98 & 3.00\\
 S20 &  34.35 & 32.58 & 1.77\\  
  			   \vspace{0.5em}
 S40 &  31.74 & 30.78 & 0.96\\
 BH10 & 35.37 & 24.03 & 11.34\\
 BH20 & 31.84 & 23.98 & 7.86\\ 
 BH40 & 28.48 & 23.88 & 4.60\\
 \hline 
\end{tabular}}
 \vspace{1em}
 \captionof{table}{ %\color{red}
 Ablation study of BM3D-AF and -MF extensions 
 on 5-frame datasets.
 \textbf{Left:} Degradation in decibels due to BM3D-AF+ variant. 
 \textbf{Right:} Deterioration due to BM3D-MF+ variant. 
 Abbreviations are as in Tables \ref{table2} and \ref{table2_1}. }
 \label{table2_ablation}
\end{table}
%--------------------------------------------------------------% 
\subsection{Perfectly Registered Datasets}
Tables \ref{table2} and \ref{table2__1} showcase 
the PSNR values of the denoised images, and Figure \ref{fig:res3}
displays the visual results after we have applied all ten 
methods. It is clear from these results that extensions of type-AF 
outperform all other techniques. They are superior to type-MF 
approaches (which is in contradiction to our conference paper
\cite{bodduna2019MultiFrame}) as we account for the change 
in the noise distribution due to temporal averaging. 

In the category-FA extensions, we directly apply the single-frame 
filters on every frame. This is a sub-optimal solution because we 
do not have enough signal on each of the frames. Techniques belonging 
to type-SF do not make use of the complete available 
information as they just consider a single reference frame.
 
In the MF and CF filters, we avoid the disadvantages of both 
FA and SF. However, they fall behind type-AF methods for two 
reasons: Firstly, we separate out temporal and spatial filtering 
in category-AF techniques. This is advantageous since we have 
noisy versions of the same original gray value in the temporal 
dimension for perfectly registered images. In the spatial 
dimensions we have noisy versions of approximately equal 
gray values in general. 
This outperforms simultaneous non-linear filtering of 
the MF and CF techniques, where we combine the information in 
all dimensions at one go. Such a strategy proves to be inferior 
even though we use a non-linear filtering in the temporal dimension 
when compared to the linear temporal averaging of category-AF filters. 
Interestingly, a similar result was observed in a single-frame 
scenario in the work of Ram et al. \cite{ram2013image}.
By adopting a simple linear filtering on a smoothly 
reordered set of pixels they could produce results almost equivalent
to the sophisticated BM3D filtering.
The reason behind such observations is that 
linear averaging of different noisy versions of 
the same pixel intensity does not create artifacts like 
a non-linear combination of dissimilar intensities does.
This is also the reason why averaging is preferred 
in electron microscopy \cite{Fr2013}.
Moreover, the linear nature of temporal averaging 
helps in computing the new standard deviation of noise 
after temporal filtering through theoretical knowledge. 
The second reason why MF and CF types fall behind category-AF 
is the following: The latter extension computes the initial 
grouping on the less noisy averaged image. 
In all the other categories we do this on the 
noisy initial images, which makes the grouping error-prone. 

The overall better performance of type-AF filters does not mean 
we can immediately reject the next best MF and CF categories. 
We must remember that we assumed AWGN noise and perfect registration.
In the first scenario, we were able to optimize the denoising ability 
of NLB-AF and BM3D-AF easily for AWGN. Its signal
independent nature helped in easier selection of filtering 
parameters which account for the change in noise 
distribution after temporal averaging. For noise of Poissonian 
type for example, AWGN elimination methods are normally combined 
with variance stabilizing tranformations 
for noise elimination. These transformations have the property 
of inducing a bias while stablizing the variance in the data. 
In another recent paper
\cite{bodduna2019MultiFramePoisson}, we evaluated the first 
four BM3D extensions in the Poissonian noise scenario and  
observed similar results as for our Gaussian noise study
\cite{bodduna2019MultiFrame}: BM3D-MF outperformed BM3D-AF. 
Apart from not accounting for the change in noise distribution 
due to temporal averaging, the above mentioned bias problem 
was also a reason behind this. We conjecture that employing more 
sophisticated stabilisation frameworks 
\cite{makitalo2011ClosedForm, 
azzari2016IterativePoisson} could help in this respect. 
The second scenario where we cannot reject methods from categories 
other than type-AF is for imperfect registrations. 
We will examine this situation in the 
upcoming section where we consider non-registered datasets.

Furthermore, BM3D-AF is superior to NLB-AF (from Tables \ref{table2}, \ref{table2__1} and Figure \ref{fig:res3}) because BM3D is a better 
single-frame denoising method than NLB for gray value images. 
We infer that the usage of the discrete 
cosine transform and the bi-orthogonal spline wavelet transform 
in the two main steps of BM3D, respectively, leads 
to superior anisotropic modeling.  
% -----------------------------------------------------------%
\subsection{Non-registered Datasets}
Tables \ref{table4}, \ref{table5} and \ref{table6} display the 
PSNR values of the denoised images while Figures \ref{fig:res4} 
and \ref{fig:res7} showcase the visual results. It can be clearly 
seen that NLB-AFand BM3D-AF outperform other approaches several 
times. However, for low amplitude noise situations NLB-CF, which 
is the current state-of-the-art method, is competitive 
with the category-AF extensions and even superior to them 
at certain occasions. 
Let us explore these results a bit further. 
For all the three datasets, we have performed experiments 
on two kinds of data: One with less number of frames and the 
other with more of them. In the latter case 
it is highly probable that there exists 
large motion between the reference frame and others which can
lead to high errors in motion estimation. Hence, if a particular 
approach is able to produce better quality results for 
a high number of frames, this indicates that it is robust 
to motion estimation errors.  
From Tables \ref{table4}, \ref{table5} and 
\ref{table6}, we can observe that CF is the only 
technique which does not even have a single instance 
where the PSNR value has decreased when more number of 
frames have been utilized. AF, MF, FA and category-SF filters 
could produce enough quality improvement for 
perfectly registered data. However, in the present non-registered 
layout we can find at least one instance for each of these 
extensions where the quality has deteriorated with an increase 
in number of frames. The only explanation behind this is the 
robustness of category-CF extensions with respect to motion. 
However, at regions where 
the motion registration is correct, the performance of AF-type
techniques is so high that they can outperform category-CF approaches
despite presence of motion estimation errors at other regions. 
Nevertheless, optical flow methods will 
continue to improve in the future. 
Thus, the philosophy of our proposed 
category-AF extensions will benefit from these advancements.

As already mentioned, the BM3D-MFO variant 
employs twice the number of patches than BM3D-MF. 
The decrease in PSNR from BM3D-MF to BM3D-MFO in Table 
\ref{table5} for high noise amplitudes 
and visual results in Figure \ref{fig:res6} 
indicate the following: The black patches in darker regions of 
the image can be eliminated using BM3D-MFO. 
However, we must use the above strategy of increasing the 
number of patches only if we encounter black patches.
Having too many them in a 3D group would  
instead give rise to an undesirable blurring. 

{In order to emphasise the critical nature of  
noise standard deviation selection as well as optical flow-based 
registration, we have performed two small ablation studies. 
Figure \ref{fig:res7_1} illustrates the importance of 
selecting the correct noise standard deviation. 
One might also argue that there is 
no need for an optical flow-based registration in category-MF 
extensions. They inherently possess a patch-based search algorithm 
which can compensate for motion within frames. However, such a  
strategy assumes a translatory motion. The optical flow approaches, 
on the other hand, are applicable for any type of motion.
Figure \ref{fig:res7_2} shows the significance 
of optical flow-based registration.}
{ %\color{red} 
In Table \ref{table2_ablation}, we present a more 
detailed ablation study for non-registered data. } 

Thus, we can draw two conclusions from our results: The latest 
robust optical flow methods are also capable of extending the best 
performing nature of type-AF filters from the perfectly registered 
layout to the non-registered scenario. Secondly, in the future we 
should concentrate on approaches which separate the 
filtering in spatial and temporal dimensions for ideal  
as well as practical situations, like BM3D-AF and NLB-AF. 

In recent years, learning-based denoising solutions have gained 
a lot of attention. In order to finish a comprehensive evaluation 
of our proposed technique, we have also compared its performance 
with a state-of-the-art neural network-based filter - 
VNLNET \cite{davy2019non, davy2020video}. 
Table \ref{table2_1} shows the PSNR 
values of this evaluation. The results show that our strategy 
outperforms VNLNET in the perfectly registered scenario and 
is competitive with it in the non-registered layout.

All the above results show that type-AF filters are among the 
best performing methods 
irrespective of whether there is any motion or not in the 
image dataset, what criteria have been used to optimize the optical 
flow, and what kind of optical flow technique has been employed. 
In future, BM3D-AF and NLB-AF can be combined with occlusion 
handling \cite{buades2016patch}, deflickering 
and sharpening \cite{maggioni2012video} strategies. 
One could also replace the present denoising and motion 
estimation techniques with better ones 
for further pushing the state-of-the-art standard. 
% -----------------------------------------------------------%

The AF-type frameworks are also the fastest among all  
extensions as they employ separable spatio-temporal filtering.
Since temporal averaging can be performed in 
real time, their net complexity is just a combination of the 
optical flow method and the 2D single-frame filter employed on 
the temporally averaged frame.  
Although all the experiments in this paper were performed using a  
CPU (Intel(R) Core(TM) i7-6700 CPU @3.4 GHz using C++ and 
OpenMP) implementation, we also have a 
GPU (NVIDIA GeForce GTX 1070 graphics card using 
ANSI C and CUDA) version of BM3D-MF.  
We have already shown that BM3D-MF encompasses the original 
single-frame BM3D algorithm mathematically.
Thus, the same GPU implementation can also be employed for BM3D-AF 
by just changing the number of frames to one and using the new 
standard deviation of noise after temporal averaging, as input. 
With such an approach, we have observed that BM3D-AF is
7.25 times faster than BM3D-MF for a 4$\times$640$\times$480 
sized dataset. It consumes just 1.82 seconds for the filtering 
process after motion compensation, despite  
employing a naive patch matching algorithm. Also,
the CPU$^2$ implemetation of BM3D-AF is over 50 times faster 
than NLB-CF, which is a current state-of-the-art technique.
% -----------------------------------------------------------%
%\subsection{Implementation Details}
%For a dataset with $L$ frames, our experiments on an 
%Intel(R) Core(TM) i7-6700 CPU @3.4 GHz machine using 
%OpenMP indicate that BM3D-MF is slightly
%more than $2L$ times slower than BM3D. 
%This is because we have two main steps, $L$ frames and a 
%comprehensive patch-similarity search across all frames.
%We have also implemented BM3D-MF on an NVIDIA 
%Quadro P5000 architecture using CUDA which takes 2.7 seconds 
%for a dataset sized $256 \times 256 \times 5$. However, we should 
%remember that this additional computational 
%cost is a trade-off for significant qualitative enhancement: 
%For a ten-frame $\sigma_\textrm{noise} = 80$ combination, 
%we attained an improvement of 28\% for the Lena dataset
%\cite{bodduna2019MultiFrame} and 26\% for the Bird House dataset, 
%over the next best single-frame denoising extensions.
% -----------------------------------------------------------%
\section{Conclusions and Outlook}
\label{sec:conclusion}
We have optimized the usage of NLB and BM3D filters 
for the multi-frame scenario. 
We can conclude from the experiments that our proposed 
following sequential process gives the best results in most 
cases: They register the images with robust 
optical flow methods, temporally average the registered noisy 
images, and then apply the single-frame filters with optimal 
parameters corresponding to the new noise distribution after 
temporal averaging. 
This is true for both NLB and BM3D, an observation which has 
surprisingly not been recognized for many years. 
This re-affirms the fact that sometimes the simpler solutions 
are the most powerful ones and can also be competitive with 
sophisticated neural network architectures.
Furthermore, we achieve this 
significant quality improvement at the cost of zero 
additional parameters and far less computational time. 
The technique also preserves a large amount 
of detail even when the images are corrupted with noise of very 
high amplitude. Thus, the category-AF extensions in combination 
with robust optical flow methods can 
be employed in practice for many multi-frame image 
processing applications.

Combining BM3D-AF and NLB-AF 
with variance stabilizing transformations, 
deflickering, sharpening and occlusion handling techniques 
will be considered in our future research. We will also use 
type-AF extensions as regularizers in PDEs for robust 
image reconstruction applications; c.f. \cite{buades2019motion, 
buades2019enhancement, BW2017, bodduna2019hough}.  
% -----------------------------------------------------------%
\par {\bf Acknowledgements.}
J.W. has received funding from the European Research Council (ERC)
under the European Union's Horizon 2020 research and innovation 
programme (grant no. 741215, ERC Advanced Grant INCOVID).

We thank Prof. Karen Egiazarian from Tampere 
University, Finland. A valuable discussion with him has helped 
to improve the evaluation part of this work. We also 
thank Dr.~Matthias Augustin and Dr. Pascal Peter 
for useful comments on a draft version of the paper.
% -----------------------------------------------------------%

%%%%% References %%%%%

\bibliography{myrefs}   % bibliography data in report.bib
\bibliographystyle{spiejour}   % makes bibtex use spiejour.bst

%%%%% Biographies of authors %%%%%

\vspace{2ex}\noindent\textbf{Kireeti Bodduna} 
received his Bachelors and Masters in Physics from the Indian 
Institute of Science Education and Research (Kolkata, India), in 2014.
After working as a high school physics teacher in Hyderabad, India, 
for about an year, he joined the 
Saarbr{\"u}cken Graduate School of Computer Science in Germany. 
Here, he first completed graduate level coursework and 
is now a Ph.D. student at the Mathematical Image Analysis 
Group, Saarland University (Saarbr{\"u}cken, Germany).   
Kireeti Bodduna is interested in developing 
traditional models for image denoising, 
super resolution, structure enhancement, motion and
depth estimation in 2D, 3D and temporal data layouts. 

\vspace{2ex}\noindent\textbf{Joachim Weickert} 
is a Professor of Mathematics and Computer Science at Saarland 
University (Saarbr{\"u}cken, Germany), where he heads the 
Mathematical Image Analysis Group. He graduated and
obtained his Ph.D. from the University of Kaiserslautern
(Germany) in 1991 and 1996. He worked as post-doctoral researcher 
at the University Hospital of Utrecht (The Netherlands) and the 
University of Copenhagen (Denmark), and as assistant professor
at the University of Mannheim (Germany). Joachim Weickert has 
developed many models and efficient algorithms for image processing 
and computer vision using partial differential equations and 
variational methods. He has served in the editorial boards of ten
international journals or book series and is Editor-in-Chief of 
the Journal of Mathematical Imaging and Vision. 
In 2010, he has received a Gottfried Wilhelm Leibniz Prize 
and in 2017 an ERC Advanced Grant for inpainting-based 
compression of visual data. 

%\vspace{1ex}
%\noindent Biographies and photographs of the other authors are not %available.

%\listoffigures
%\listoftables

\end{spacing}
\end{document}